\documentclass[a4paper,11pt]{article}
\usepackage{jcappub} 
\usepackage{lineno}

\newcommand{\UDF}{\Omega_\textrm{UDF}}

\newcommand{\sixxtwo}{$6\times2$\,pt}
\newcommand{\lcdm}{$\Lambda\textrm{CDM}$}
\newcommand{\SO}{{\em Simons Observatory}}

\title{Unified dark fluid with null sound speed as an alternative to phantom dark energy}

\author[1]{Raphaël Kou\note{Corresponding author.}}
\author{and Antony Lewis}
\affiliation{Department of Physics \& Astronomy, University of Sussex, Brighton BN1 9QH, UK}

\emailAdd{r.kou@sussex.ac.uk}
\emailAdd{antony@cosmologist.info}
\abstract{
Recent BAO measurements from DESI, when combined with CMB and supernovae data, suggest evolving dark energy and in particular point to a possible phantom regime, with an equation of state parameter $w<-1$. This behaviour is theoretically problematic, as it violates the null energy condition and typically leads to instabilities in the perturbations.

We explore an alternative phenomenological way to model dark matter and dark energy based on a unified dark fluid (UDF). By construction, our model reproduces the same background expansion history as DESI's best-fit using the Chevallier–Polarski–Linder (CPL) parametrization, but assumes a vanishing rest-frame sound speed and no anisotropic stress. This simple prescription ensures a consistent and physical treatment of perturbations and, in our case, the use of a unified dark sector avoids phantom behaviour. We model CMB, large-scale structure, and redshift-space distortion observables, and find mostly small differences with CPL, suggesting that while stage IV CMB and galaxy surveys will be able to test these models, achieving a decisive distinction between them may prove challenging on linear scales. At the non-linear level, we study spherical collapse in the UDF and show that within this framework, structure formation proceeds very similarly to standard scenarios.

Using Planck, DESI BAO DR2, and DES Y5 supernovae data, we demonstrate that this simple UDF model fits current observations nearly as well as CPL, while treating perturbations consistently. Because most cosmological observations are not sensitive to how the dark sector is split, the unified framework can also approximate the phenomenology of interacting dark energy–dark matter scenarios or evolving dark matter, making it a general way to model the data, at least as long as the dark components have a vanishing sound speed, which is the most distinctive feature of our analysis. Our results highlight that a unified dark fluid with evolving equation of state and null sound speed is sufficient to pass current constraints without invoking a phantom component.}
\begin{document}
\maketitle
\flushbottom

\section{Introduction}
\label{sec:intro}

The nature of the dark energy responsible for cosmic acceleration remains uncertain. While the cosmological constant is still consistent with many observations, recent Baryon Acoustic Oscillations (BAO) measurements from the DESI collaboration~\cite{DESI:2025zgx}, in combination with Cosmic Microwave Background (CMB) and supernovae data, have shown interesting hints for evolving dark energy. Modelling the dark energy equation of state through the  Chevallier-Polarski-Linder (CPL) parametrization~\cite{Chevallier:2000qy,Linder:2002et}
\begin{align}
    w(a)=w_0+w_a(1-a),
\end{align}
they found a preference for phantom crossing dark energy~\cite{Caldwell:1999ew}, where the dark energy equation of state, being greater than $-1$ at low redshift, becomes lower than $-1$ at high redshift. This hint for phantom crossing is not only found with the CPL parametrization, but is also preferred when using other analytical parametrizations or when reconstructing the equation of state with Gaussian processes, redshift bins or splines~\cite{Giare:2024gpk,DESI:2024aqx,Reboucas:2024smm,Wolf:2025jlc,DESI:2025fii}. More quantitatively, \cite{DESI:2025fii}~found that the CPL parametrization with phantom crossing improved the fit by $\Delta\chi^2 = 7.8$--$10.5$ for combinations of BAO, CMB, and supernovae data, with the exact value depending on the supernova sample used, compared to the non-phantom crossing case. While this improvement is not negligible, it only represents a mild preference for phantom crossing models.

When dark energy is modelled as a single scalar field, this preference for a phantom crossing is however problematic for several reasons~\cite{Carroll:2003st,Vikman:2004dc}. Firstly, the density of a phantom dark energy increases over time, and thus violates the Null Energy Condition (NEC)~\cite{Sen:2007ep,Lewis:2024cqj}, which would have dramatic consequences for our understanding of physics. Secondly, even though the preference for phantom dark energy relies mainly on observations of the background expansion of the Universe, describing the perturbations in this regime is very tricky as phantom dark energy tends to induce singularities in the perturbations. It must be noted, however, that more complex dark energy models, such as the ones involving multiple scalar fields, can effectively cross the phantom divide, without suffering from those issues~\cite[e.g.,][]{Hu:2004kh,Feng:2004ad,Cai:2025mas}.

Background expansion measurements, such as with BAO and supernovae, are only sensitive to the evolution of the total energy density of the Universe as determined by the Friedmann–Lemaître equations. They do not depend on how this density is divided among the various components. In particular, such observations cannot distinguish how the total dark sector is split between dark matter and dark energy. This indeterminacy leaves us free to partition the dark sector in different ways, a freedom known as the ``dark degeneracy''~\cite{Wasserman:2002gb,Kunz:2007rk,Kunz:2009yx,2011PhRvD..84f3519S,Aviles:2011ak,Carneiro:2014uua,vonMarttens:2019ixw}. Combining dark matter and dark energy in a unified dark fluid~\cite[e.g.,][]{Makler:2002jv,Reis:2004hm,Brandenberger:2019pju,2019IJMPD..2850110E,Frion:2023xwq,Wang:2024rus,Su:2025ntt} is therefore equivalent at the background level, and avoids the need for phantom components, as we shall see in section~\ref{sec:theory}. This approach can therefore reproduce the background expansion preferred by recent measurements, while describing the perturbations in a physical way and without violating the NEC.

We present our theoretical model in section~\ref{sec:theory}, before investigating its effect on CMB and Large-Scale Structure (LSS) observations in section~\ref{sec:observables}. We constrain this model using recent data and compare it to the standard CPL model in section~\ref{sec:data}. Finding very similar results between our unified model and CPL, we test whether those two models will be distinguishable with stage IV data in section~\ref{sec:forecast}. Finally, in section~\ref{sec:non_linear} we study the theoretical predictions of the spherical collapse with our model, to test whether it can reproduce small-scale structure formation.

\section{Theoretical model}\label{sec:theory}

In this work, we describe the density content of the Universe through different components, namely radiation ($\Omega_r$), baryons ($\Omega_b$), massive neutrinos ($\Omega_\nu$) and a Unified Dark Fluid (UDF) ($\UDF$). The Friedmann-Lemaître equation therefore follows
\begin{align}
    H^2(a)=H_0^2\left[\Omega_ra^{-4}+\Omega_ba^{-3}+\Omega_\nu(a)+\UDF(a)\right],
    \label{eq:Friedmann}
\end{align}
where $H(a)$ is the Hubble rate at a scale factor $a$ and $H_0$ is the Hubble rate today ($a = 1$). The UDF modelling relies on the generalized dark matter approach~\cite{Hu:1998kj}, in which a fluid can be described at the background and perturbation levels by setting the fluid's equation of state, sound speed, and anisotropic stress. In particular, in order to reproduce the background expansion described in the CPL parametrization, we impose

\begin{align}\label{eq:w_of_a}
    w_\textrm{UDF}(a)&=\frac{\Omega_\textrm{de}(a)w_\textrm{de}(a)+\Omega_c(a)w_c(a)}{\Omega_\textrm{de}(a)+\Omega_c(a)} \\
    &=\frac{\Omega_\textrm{de}e^{-3w_a(1-a)}a^{-3(1+w_0+w_a)}(w_0+w_a(1-a))}{\Omega_\textrm{de}e^{-3w_a(1-a)}a^{-3(1+w_0+w_a)}+\Omega_ca^{-3}},
\end{align}

where $(\Omega_\textrm{de},w_\textrm{de})$ and $(\Omega_c,w_c=0)$ are the dark energy and cold dark matter density and equation of state parameters in the non-unified case, using CPL to parametrize the dark energy evolution. Note that the parameters $\Omega_c$, $w_0$ and $w_a$ are only used here to parametrize the equation of state of the UDF, and are not used in any other equation. We also clarify that for convenience, whenever we refer to CPL, we implicitly think about the non-unified model in which dark matter is assumed to be cold, and dark energy is parametrized with the CPL model. We hence loosely use the acronym ``CPL'' instead of ``CPL+CDM''.

\begin{figure}[htbp]
\centering
\includegraphics[width=0.8\textwidth]{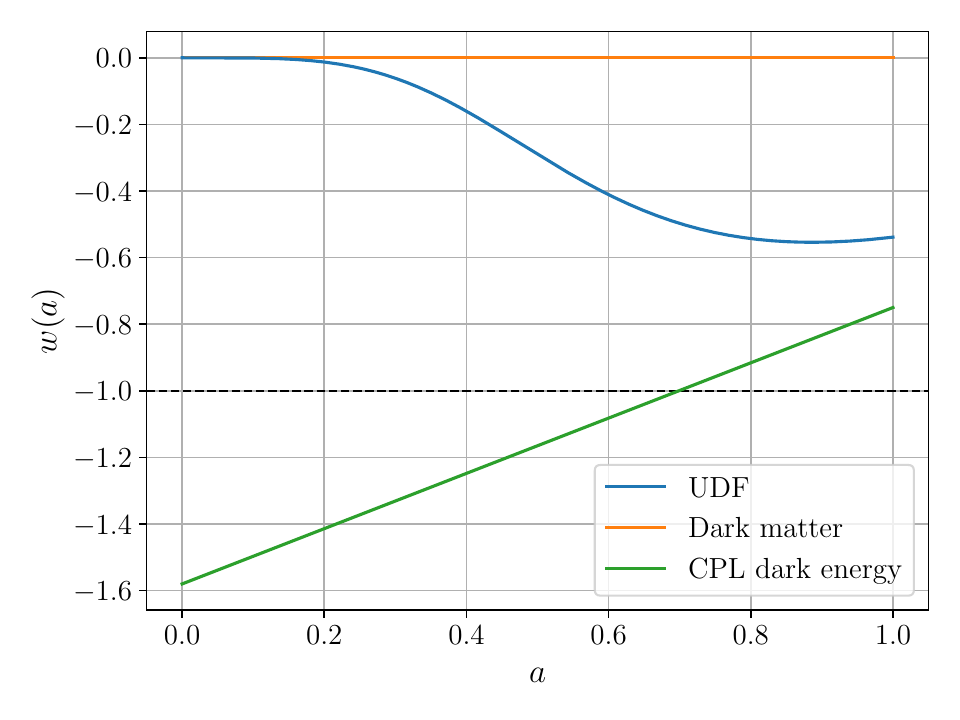}
\caption{UDF equation of state evolution with scale factor. For comparison, we also show the dark matter and dark energy equations of state as a function of scale factor. The dark energy equation of state corresponds to the best-fit to DESI BAO DR2, Planck CMB and DESY5 supernovae obtained in the CPL parametrization.\label{fig:w_a}}
\end{figure}

Figure~\ref{fig:w_a} shows the evolution of the equation of state of the UDF as a function of scale factor, using the best-fit values for $w_0$, $w_a$ and $\Omega_c$ obtained using Planck (PR4 NPIPE~\cite{Rosenberg:2022sdy} at high multipoles, 2018~\cite{Planck:2019nip} at low multipoles), DESI BAO DR2~\cite{DESI:2025zgx,DESI:2025zpo} and supernovae from the Dark Energy Survey Year 5 (DESY5)~\cite{DES:2024jxu}. As it can be seen, the equation of state is null at low scale factors, where dark matter dominates (in the non-unified case), and takes negative values at higher scale factors when dark energy becomes significant (still in the non-unified case). While this parametrization of the UDF provides the exact same background expansion as the CPL model by construction, it does not involve any phantom crossing, showing that the hint for phantom dark energy depends on how the dark sector is split. In particular, interacting dark energy and dark matter, evolving dark matter, or truly unified dark fluid as modelled in this work could be alternative ways of describing the background expansion, without requiring any phantom dark energy~(see e.g.~\cite{Khoury:2025txd,Chakraborty:2025syu,Guedezounme:2025wav,Chen:2025wwn,Braglia:2025gdo,Su:2025ntt}).

At the perturbation level, however, setting such an equation of state is not enough to ensure that the dark degeneracy holds. Indeed, the evolution of the perturbations depends on the sound speed and anisotropic stress of the fluid, two quantities that must be set in the generalized dark matter approach in order to specify how the fluid is perturbed. By choosing the right redshift and scale dependence for those two quantities, it is then possible to maintain the dark degeneracy for linear perturbations. However, while this approach would allow reproducing exactly the evolution of both the background and the linear perturbations in the CPL parametrization with a unified model, we show here that such a fine-tuning is not required to pass most existing cosmological constraints. In particular, we neglect the anisotropic stress, and use a constant, null sound speed, a case that has already been studied in the literature for non-unified models (see e.g.~\cite{Creminelli:2008wc,Creminelli:2009mu,Sefusatti:2011cm,DAmico:2011ldy,Anselmi:2011ef,Kunz:2015oqa,Lewandowski:2016yce,Hassani:2020buk,DAmico:2020tty,Batista:2021uhb,Lu:2025gki}). Using a close-to-null sound speed is required at high redshift in order to pass CMB constraints when dark matter dominates, but not necessarily at low redshift, a priori. The most interesting feature of this work is therefore to show that many current cosmological observables can be reproduced with a simple model with evolving equation of state and a null sound speed. For clarity, we emphasize that while our UDF is fully equivalent to CPL at the background level, this is not the case at the perturbation level.

In particular, when dark energy is described as a quintessence field, its rest-frame sound speed is equal to $1$, which would lead to a non-zero sound speed for the combined dark matter-dark energy fluid. If the sound speed of dark energy were $0$ however, the two models would be equivalent if dark energy was also comoving with dark matter. More generally, our UDF model approximates well any split into different subcomponents (with evolving equations of state, interactions, etc.) with null sound speeds. Interestingly, it has been shown~\cite{Creminelli:2008wc} that a single-field dark energy model could cross the phantom divide without any pathology in case its sound speed is exactly null. Dark energy models with equation of states below $-1$ which do not cross the phantom divide would require negative, and still very low sound speeds such that $-\hat{c}^2_s \lesssim 10^{-30}$~\cite{Creminelli:2008wc}. Our unified model could hence be interpreted as a unification of comoving cold dark matter and evolving dark energy with null sound speed. Even though such a unification would not be theoretically required, if both dark energy and dark matter had a vanishing sound speed and were comoving, it would seem very natural to think of them as a single component. In this case, there is no phantom divide crossing to consider, so there would be no need to impose an exactly null sound speed specifically for that purpose.

We implement our model in a modified version of CAMB~\cite{Lewis:1999bs,Howlett12}, such that the background expansion is described through equation~\ref{eq:Friedmann}, and the perturbation equations for the UDF are given in the synchronous gauge by~\cite{Weller:2003hw}
\begin{align}
    \dot{\delta}+3\mathcal{H}(\hat{c}_s^2-w)(\delta+3\mathcal{H}(1+w)v/k)+(1+w)kv+3\mathcal{H}\dot{w}v/k &=-(1+w)\frac{\dot{h}_L}{2} \\
    \dot{v} +\mathcal{H}(1-3\hat{c}_s^2)v &= \frac{k\hat{c}_s^2\delta}{1+w},
\end{align}
where $\delta$ is the fluid's overdensity, $v$ is its velocity perturbation, $\hat{c}_s^2$ is its rest-frame sound speed, $\mathcal{H}=aH$ is the comoving Hubble parameter, $h_L$ is the trace part of the metric perturbation and time derivatives (over-dots) are taken with respect to conformal time. Since we set the fluid's rest-frame sound speed $\hat{c}_s$ to $0$, those equations become
\begin{align}
    \label{eq:perturbation_density}
    \dot{\delta}-3\mathcal{H}w(\delta+3\mathcal{H}(1+w)v/k)+(1+w)kv+3\mathcal{H}\dot{w}v/k &=-(1+w)\frac{\dot{h}_L}{2} \\
    \label{eq:perturbation_velocity}
    \dot{v} +\mathcal{H}v &=0.
\end{align}
We use adiabatic initial conditions such that
\begin{align}
    \delta_{\textrm{UDF},0} &= \delta_{\textrm{b},0} \\
    v_{\textrm{UDF},0}&=0,
\end{align}
with $\delta_{\textrm{b},0}$ the initial baryon overdensity. As can be seen from equation~\ref{eq:perturbation_velocity}, initializing the synchronous-gauge velocity perturbation of the UDF to zero ensures it remains zero at all times. We now examine whether this simple yet physically consistent UDF model can reproduce key cosmological observables. In particular, we compare predictions for the CMB, matter power spectrum, lensing observables, and redshift-space distortions.

\section{Effect on observables}\label{sec:observables}
In this section, we carefully look at the effect of the UDF on cosmological observables and compare the predictions of this model to the ones of CPL. In all cases, we compare the predictions of our UDF model to CPL, using the best-fit values obtained by fitting the CPL model to the combination of Planck (NPIPE at high multipoles, 2018 at low multipoles), DESI BAO DR2 and DESY5. We hence use the Hubble constant $H_0=66.71$ ($h=0.6671$ denotes the reduced Hubble parameter), the cold dark matter density $\Omega_ch^2=0.1192$, the baryon density $\Omega_bh^2=0.0222$, the optical depth $\tau=0.0520$, the amplitude of primordial scalar fluctuations $A_s=2.08\cdot10^{-9}$, the spectral index $n_s=0.964$, the CPL parameters $w_0=-0.75$ and $w_a=-0.83$. All predictions are made using linear perturbation theory as we will only consider the non-linear regime in section~\ref{sec:non_linear}. In the case of CPL, the linear theoretical modelling is done using the parameterized post-Friedmann (PPF)~\cite{Hu:2007pj,Fang:2008sn} approach implemented in CAMB.

\subsection{CMB}
\begin{figure}[htbp]
\centering
\includegraphics[width=\textwidth]{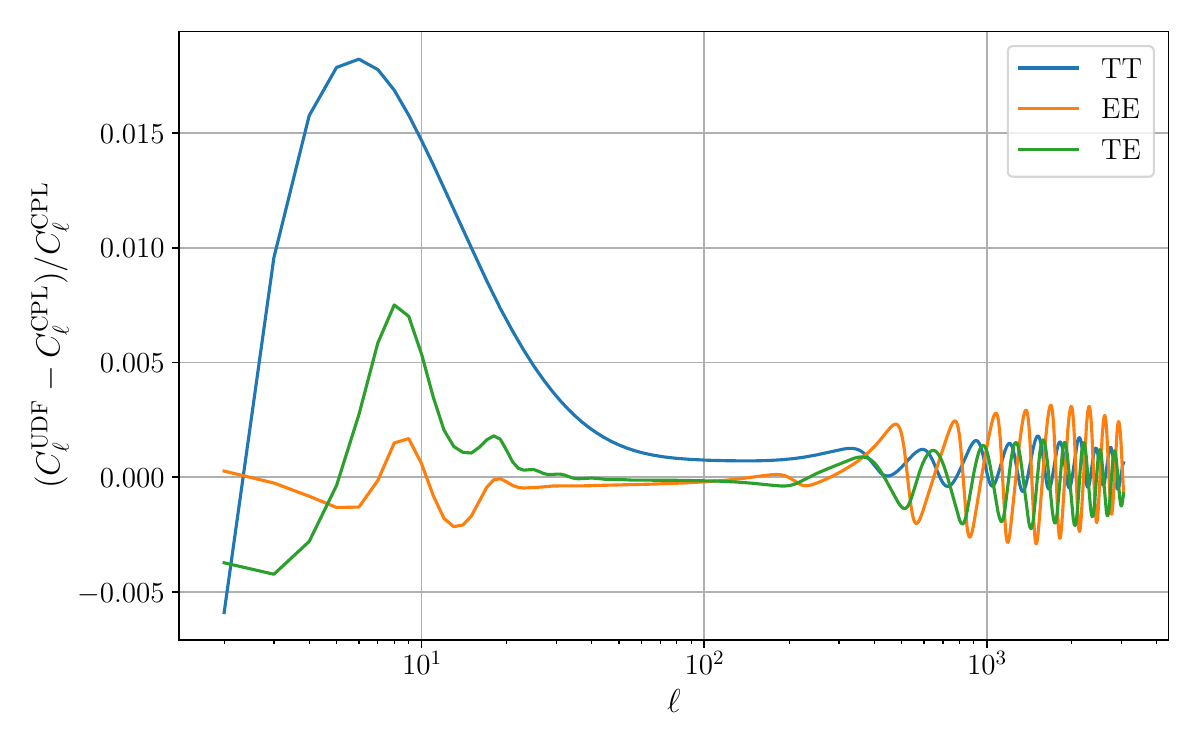}
\caption{Relative difference between CMB power spectra prediction in UDF and CPL with the fiducial values stated in the text. The relative difference is mostly subpercent, apart from the ISW effect in the temperature power spectrum. Note that in the TE case, the normalization is not $C_\ell^{TE}$ but rather $\sqrt{C_\ell^{TT}C_\ell^{EE}}$.}\label{fig:CMB}
\end{figure}

Figure~\ref{fig:CMB} shows the relative difference in the CMB temperature and E-mode polarization power spectra up to $\ell=3000$. Apart from a small difference in the Integrated Sachs-Wolfe (ISW) effect on large scales, the relative difference is always well below the percent level. The fact that the relative differences are so small is due to the fact that the two models are very much equivalent at high redshift, where the impact of dark energy is completely negligible as oscillations in the primordial plasma do not depend at all on late time physics. The differences between our UDF model and CPL lie in how perturbations grow at very low redshift, which could therefore impact the lensing and the late-time ISW. As already mentioned, the latter effect can be seen and induces a difference of about $2\%$ on large scales, which might lead to a slight shift in the parameter inference. However, that would not be enough to discard the UDF model, even for future CMB experiments, as this difference is much below the cosmic variance level on the relevant scales.

\subsection{Matter power spectrum}
We then look at the matter power spectrum. However, there is an ambiguity about what we call ``matter'' in the UDF model, and about what galaxies trace. Indeed, the matter power spectrum is usually defined as the $2$-pt function of the matter overdensity
\begin{align}
    \langle\delta_m(k,z)\delta_m^*(k',z)\rangle = (2\pi)^3\delta_D(k-k')P_m(k,z),
\end{align}
with $\delta_m=\frac{\delta \rho_c+\delta\rho_b}{\rho_c+\rho_b}$. In the UDF case, we instead define the power spectrum using the overdensity of the total fluid, including the UDF and baryon contributions such that $\delta_\textrm{tot}=\frac{\delta \rho_\textrm{UDF}+\delta\rho_b}{\rho_\textrm{UDF}+\rho_b}$. The matter power spectrum we obtain using this definition is much lower than the one obtained using the more standard definition, which excludes dark energy density from the denominator. This is not an observational issue, as the matter power spectrum is not observable per se, and as the definition of the galaxy bias must be changed consistently. Indeed, we now make galaxies tracers of baryons and UDF field so that the galaxy bias $b_g^\textrm{tot}$ is defined such that $\delta_g=b_g^\textrm{tot}\delta_\textrm{tot}$, where $\delta_g$ is the galaxy overdensity, instead of the usual $\delta_g=b_g^m\delta_m$. However, this means that when observing how galaxies cluster, the galaxy bias $b_g^\textrm{tot}$ will tend, at low redshift, to be much higher than $b_g^m$ in order to compensate for the lower matter power spectrum. This would also impact the way galaxy bias evolves with redshift, as the dark energy contribution becomes weaker at higher redshift (in the non-unified case). Similarly, this means that the value of $\sigma_8$ in this model is significantly affected by this new definition. In fact, we get $\sigma_8^\textrm{tot}=0.26$ instead of $\sigma_8=0.80$ with the fiducial values we use.
\begin{figure}[htbp]
\centering
\includegraphics[width=\textwidth]{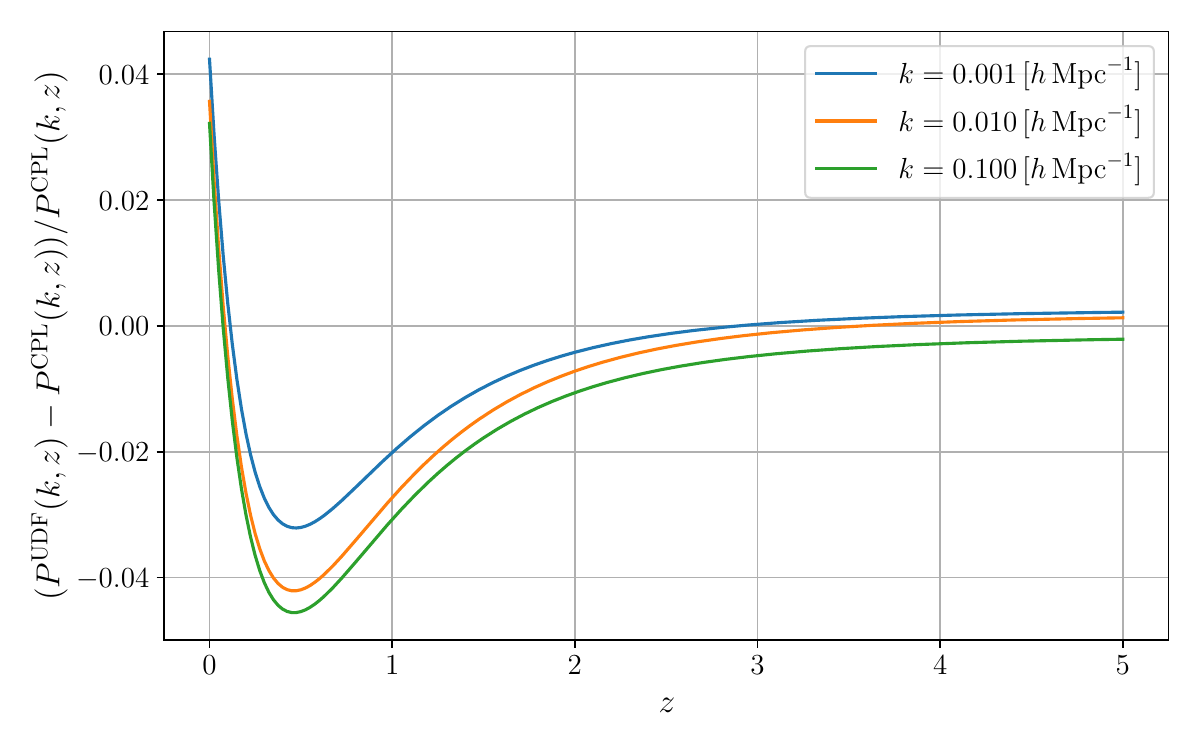}
\caption{Relative difference in the matter power spectrum with the fiducial values stated in the text. The matter power spectrum is defined with respect to the baryons + UDF overdensity in the UDF case, and baryons + dark matter + dark energy in the CPL case. This definition differs from the usual convention that excludes dark energy. Consequently, the galaxy bias would be redefined when fitting galaxy power spectra.
\label{fig:matter_power_spectrum}}
\end{figure}

Figure~\ref{fig:matter_power_spectrum} hence shows the relative difference in the matter power spectrum, using the definition we have just discussed. As can be seen, the relative difference does not depend much on scale, but evolves with redshift. The effect is most important at redshifts $z=0$ and $z\sim0.5$, and reaches $4-5\%$. At high redshift, there is almost no  difference, as the two models are almost equivalent (because the influence of dark energy is negligible). The slight scale sensitivity that can be seen on figure~\ref{fig:matter_power_spectrum} comes from the fact that the growth of structure in CPL is scale-dependent when accounting for dark energy perturbations with the PPF model (unlike in UDF, where the only scale dependence is due to the effect of baryons and massive neutrinos).

\subsection[\texorpdfstring{$6\times2$\,pt analysis}{6x2pt analysis}]%
           {\boldmath $6\times2$\,pt\textrm{ analysis}}\label{sec:6x2}
We now consider adding LSS data, namely the reconstructed CMB lensing power spectrum, the (angular) galaxy power spectrum, the galaxy shear power spectrum, and their cross-correlations, which is also sometimes called the \sixxtwo\ analysis. 

Lensing observables, either of the CMB or background galaxies, are defined as the projection over the line-of-sight of the Weyl potential whose definition is non-ambiguous and does not depend on what galaxies trace. For the fiducial galaxy distribution, required to compute galaxy and shear power spectra, we consider a Euclid-like~\cite{Euclid:2024yrr} experiment. We show the relative differences for the lowest, middle and highest redshift bins following Euclid forecasts~\cite{Euclid:2019clj}. The full galaxy distribution is hence modelled as
\begin{align}
    n(z) \propto \left(\frac{z}{z_0}\right)^2\exp{\left[-\left(\frac{z}{z_0}\right)^{3/2}\right]},
\end{align}
with $z_0=0.9/\sqrt{2}$, and is then convolved with the photometric redshift error described in~\cite{Euclid:2019clj}, which allows us to compute the galaxy distribution in $10$ redshift bins.

\begin{figure}[htbp]
\centering
\includegraphics[width=\textwidth]{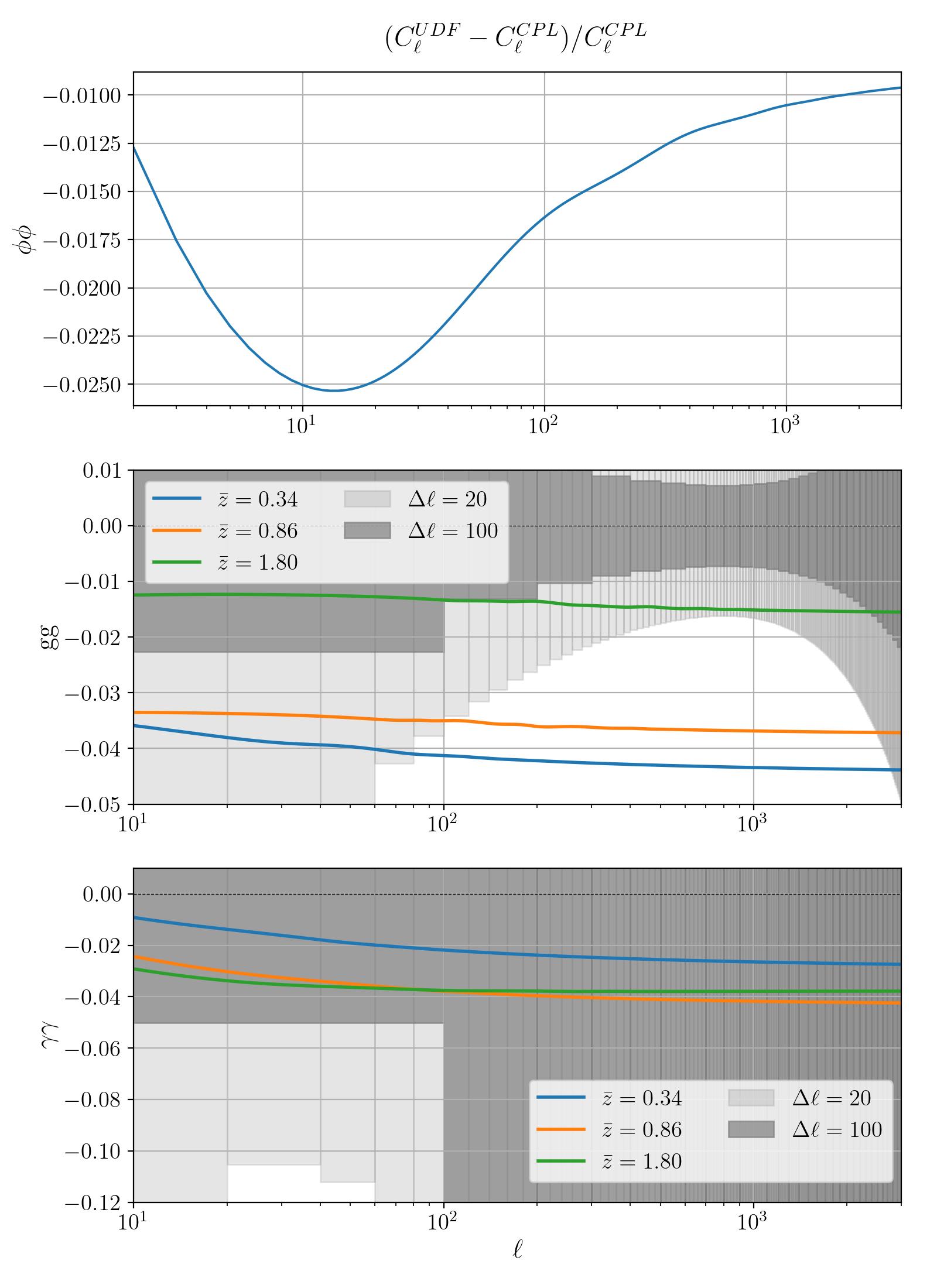}
\caption{Relative difference between CMB lensing (top panel), galaxy (middle panel) and shear (low panel) power spectra with the fiducial values stated in the text. In the latter two cases, the labels indicate the mean redshift of bins we considered. Shaded areas show the 1-$\sigma$ uncertainty that is expected for an Euclid-like experiment, using equally spaced multipole bins with width $\Delta\ell=20$ or $\Delta\ell=100$. \label{fig:6x2pt}}
\end{figure}

Figure~\ref{fig:6x2pt} shows the relative difference between CMB lensing (top panel), galaxy (middle panel) and shear (lower panel) power spectra. As mentioned before, for the latter two cases, we show power spectra corresponding to the lowest, middle and highest redshift bins of an Euclid-like experiment, indicating the mean redshift of each bin in the labels. We show relative differences over a broad range of multipoles, but since these predictions use linear perturbation theory, they are only valid over a more limited range. In practice, many of these multipoles could not be used unless the non-linear regime is also modelled. The relative difference in the CMB lensing power spectrum between CPL and UDF predictions is not more than a few percent; the CMB lensing kernel is mostly sensitive to high redshifts, where differences in the matter power spectrum are lower, as we saw on figure~\ref{fig:matter_power_spectrum}. The difference is largest at very low multipoles, which, for fixed $k$ scale, correspond to low redshifts. When looking at galaxy or shear power spectra, we find a stronger impact as those observables are sensitive to lower redshifts. In order to quantify how significant those differences are, we also show the typical expected uncertainties that an Euclid-like experiment should obtain, using a Gaussian covariance
\begin{align}
    \sigma(C_\ell)=(C_\ell+N_\ell)\sqrt{\frac{2}{(2\ell+1)\Delta\ell f_\textrm{sky}}},
\end{align}
where we use a sky coverage $f_\textrm{sky}=36\%$ and two different multipole binning schemes, equally spaced with width $\Delta\ell=20$ or $\Delta\ell=100$. The noise terms are given by $N_\ell=\frac{1}{\bar{n}_g}$ and $N_\ell=\frac{\epsilon^2}{\bar{n}_g}$ for galaxy clustering and weak lensing, respectively, with $\bar{n}_g=3\,\textrm{arcmin}^{-2}$ and $\epsilon=0.26$~\cite{Euclid:2019wjj}. The 1-$\sigma$ uncertainties are hence shown on Figure~\ref{fig:6x2pt} for the two binning schemes. Even though the relative differences are quite significant compared to the uncertainties for the galaxy power spectra, distinguishing UDF and CPL will still be challenging due to the degeneracies between parameters, as we will see in section~\ref{sec:forecast}. These differences are nonetheless some of the clearest distinct signatures of the two models and remain one of the most promising ways of distinguishing them. In particular, it will be required to constrain the galaxy bias well, as it would be able to absorb the difference we see in the galaxy power spectra otherwise. As a result, using cross-correlations between galaxy clustering and lensing measurements (galaxy shear, CMB lensing or both) will be very useful to calibrate the bias and hence to disentangle the two models.
\subsection{RSDs}
Finally, we also study the impact of this model on Redshift Space Distortions (RSD), as this probe is sensitive to peculiar velocities, and one could think that the UDF model would induce more important differences in velocity than density perturbations. We focus mainly on measurements of $f\sigma_8(z)$, where $f=\frac{d\ln{D}}{d\ln{a}}$ is the growth rate and $D$ is the growth factor such that $\delta(a)=D(a)\delta_0$, in linear perturbation theory. The evolution of the density of the total (UDF and baryons) fluid in the Newtonian gauge in linear theory is given by
\begin{align}
    \dot{\delta}_\textrm{tot}=-(1+w_\textrm{tot})(\theta_\textrm{tot}-\dot{\phi})-3\mathcal{H}\left[\dot{w}_\textrm{tot}-3\mathcal{H}(1+w_\textrm{tot})w_\textrm{tot}\right]\frac{\theta_\textrm{tot}}{k^2}+3\mathcal{H}w_\textrm{tot}\delta_\textrm{tot},
\end{align}
where $\theta_\textrm{tot}$ is the divergence of the velocity perturbation and $\phi$ is the spatial Newtonian metric potential. This expression assumes $\hat{c}_{s,\textrm{tot}}^2=0$, an approximation since the UDF and baryons do not necessarily share the same rest frame and their velocities should in principle be treated separately. We find that this simplification is excellent, as peculiar velocities remain small. Strictly speaking, baryons also have a non-zero sound speed, but this effect is relevant only on scales close to the Jeans length and is negligible on the large, linear scales of interest here.
Neglecting $\dot{\phi}$ and using $\dot{\delta}_\textrm{tot}=\mathcal{H}\delta_\textrm{tot}\frac{d\ln{D_\textrm{tot}}}{d\ln{a}}$ and $\dot{w}_\textrm{tot}=\mathcal{H}a\frac{dw_\textrm{tot}}{da}$, we get
\begin{align}
    \frac{\mathcal{H}}{1+w_\textrm{tot}+\frac{3\mathcal{H}^2}{k^2}(a\frac{dw_\textrm{tot}}{da}-3w_\textrm{tot}(1+w_\textrm{tot}))}\left[\frac{d\ln{D_\textrm{tot}}}{d\ln{a}}-3w_\textrm{tot}\right]\delta_\textrm{tot}+\theta_\textrm{tot}=0.
\end{align}
We can hence define an effective growth factor $f_\textrm{eff}$
\begin{align}
    f_\textrm{eff} = \frac{1}{1+w_\textrm{tot}+\frac{3\mathcal{H}^2}{k^2}(a\frac{dw_\textrm{tot}}{da}-3w_\textrm{tot}(1+w_\textrm{tot}))}\left[\frac{d\ln{D_\textrm{tot}}}{d\ln{a}}-3w_\textrm{tot}\right],
\end{align}
such that 
\begin{align}\label{eq:velocity_density}
    \theta_{\textrm{tot},u}=-f_\textrm{eff}\delta_\textrm{tot},
\end{align}
where $\theta_{\textrm{tot},u}$ follows the definition of~\cite{Percival:2008sh} where velocities are expressed in units of the Hubble velocity ($\theta_{\textrm{tot},u}=\theta_\textrm{tot}/\mathcal{H}$). At the fiducial cosmology we use, we find that for $k>10^{-3}h\,\textrm{Mpc}^{-1}$, $1+w_\textrm{tot} \gg \frac{3\mathcal{H}^2}{k^2}(a\frac{dw_\textrm{tot}}{da}-3w_\textrm{tot}(1+w_\textrm{tot}))$, so we can then approximate
\begin{align}
    f_\textrm{eff} \simeq \frac{1}{1+w_\textrm{tot}}\left[\frac{d\ln{D_\textrm{tot}}}{d\ln{a}}-3w_\textrm{tot}\right].
\end{align}
Using this equation, the derivations of~\cite{Percival:2008sh} remain valid as long as $f_\textrm{eff}$ and total quantities (density, equation of state, etc.) are used consistently. As a result, RSDs provide a measurement of $(f_\textrm{eff}\sigma_8^\textrm{tot})(z)$, theoretically given by
\begin{align}\label{eq:rsd}
    (f_\textrm{eff}\sigma_8^\textrm{tot})(z)=\frac{D_\textrm{tot}\sigma_{8,0}^\textrm{tot}}{1+w_\textrm{tot}}\left[\frac{d\ln{D_\textrm{tot}}}{d\ln{a}}-3w_\textrm{tot}\right],
\end{align}
where $\sigma_{8,0}^\textrm{tot}$ is the value of $\sigma_8^\textrm{tot}$ at $z=0$. As in the case of galaxy power spectra we mentioned earlier, this expression also assumes that galaxies are tracing the total (baryon and UDF) fluid.
\begin{figure}[htbp]
\centering
\includegraphics[width=0.75\textwidth]{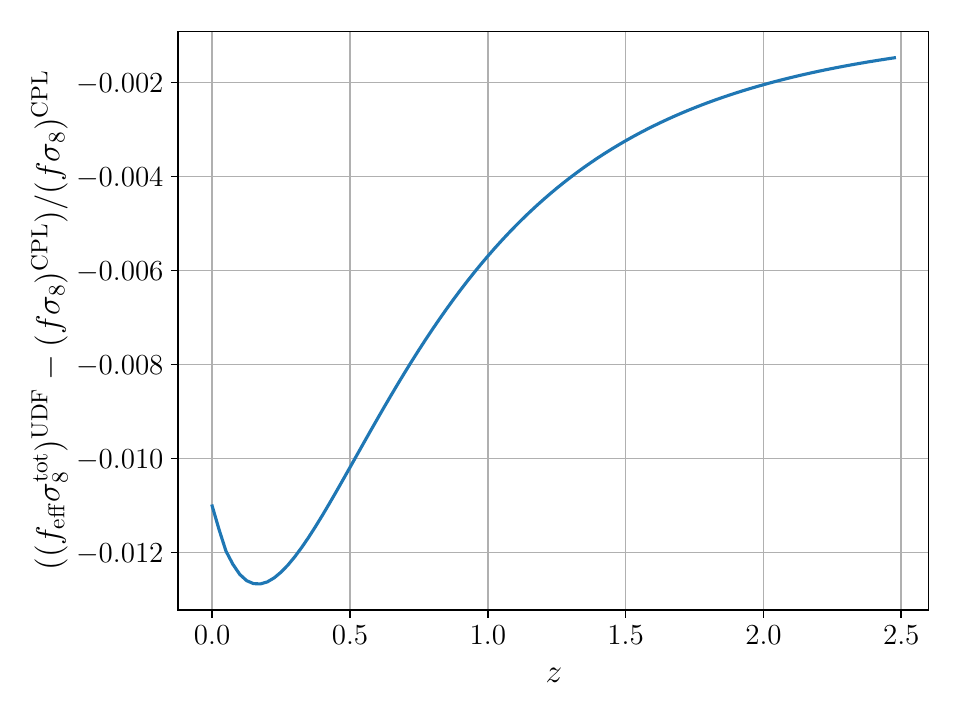}
\caption{Relative difference between $(f_\textrm{eff}\sigma_8^\textrm{tot})(z)$ in UDF and $(f\sigma_8)(z)$ in CPL as a function of redshift for our fiducial cosmology. \label{fig:rsd}}
\end{figure}

In the \lcdm\ case, we can show that choosing to model galaxies as tracers of the total fluid (baryons and UDF, where UDF hence combines dark matter and the cosmological constant) has no consequence on theoretical predictions for RSDs. Indeed, in the \lcdm\ case, we have 
\begin{align}
    1+w_\textrm{tot}(a)=\frac{\rho_m(a)}{\rho_\textrm{tot}(a)}=\Omega_m(a),
\end{align}
and for the linear perturbations,
\begin{align}
    \delta_\textrm{tot}(a) &=\Omega_m(a)\delta_m(a) \\
    D_\textrm{tot}(a) &=D_m(a)\frac{\Omega_m(a)}{\Omega_m} \\
    \sigma_{8,0}^\textrm{tot}&=\Omega_m\sigma_{8},
\end{align}
where $D_m$ is the matter growth factor and $\sigma_8$ is the standard matter variance at $z=0$. Then, $\frac{d\ln{D_\textrm{tot}}}{d\ln{a}}=\frac{d\ln{D_m}}{d\ln{a}}+\frac{d\ln{\Omega_m(a)}}{d\ln{a}}$ and it can be shown that $\frac{d\ln{\Omega_m(a)}}{d\ln{a}}=-3(1-\Omega_m)=3w_\textrm{tot}$. Replacing $D_\textrm{tot}$, $\frac{d\ln{D_\textrm{tot}}}{d\ln{a}}$ and $\sigma_{8,0}^\textrm{tot}$ in equation~\ref{eq:rsd}, we get
\begin{align}
    (f_\textrm{eff}\sigma_8^\textrm{tot})(z)=D_m(a)\sigma_{8,0}\frac{d\ln{D_m}}{d\ln{a}}=(f\sigma_8)(z).
\end{align}
We hence showed that our effective relation (equation~\ref{eq:rsd}) matches the standard result in the \lcdm\ case, and proved that, within \lcdm, RSD predictions are not impacted by our change of convention.

Returning to our fiducial model, we show on figure~\ref{fig:rsd} the relative difference between $(f_\textrm{eff}\sigma_8^\textrm{tot})(z)$ in UDF and the standard $(f\sigma_8)(z)$ in CPL as a function of redshift. We find very small differences, slightly above one percent at low redshift. Such small differences would be very difficult to distinguish, even for future experiments. As a result, among the cosmological observables that we considered, combining galaxy clustering and lensing measurements remains the most promising way of distinguishing the two models.

\section{Comparison with data and discussion}\label{sec:data}
Even though the differences on most observables between our UDF and CPL are small, we checked whether they could have an impact on the goodness of fit with current data. In particular, we considered Planck PR4 temperature and polarization power spectra, DESI BAO DR2 and DES Y5 supernovae. Finally, we also included Planck PR4 lensing~\cite{Carron:2022eyg} power spectrum, as we found that the UDF could introduce a few-percent relative differences compared to CPL, especially at large scales. We ran Monte Carlo Markov Chains (MCMC) using the Cobaya~\cite{2019ascl.soft10019T,Torrado:2020dgo} sampler, and minimized the likelihood to compute the best-fit values. Table~\ref{tab:best_fit} shows those best-fit likelihoods for the different combinations of data we looked at, both in the UDF and CPL models.

\begin{table}[htbp]
\centering
\begin{tabular}{c|c|c}
\hline
Data&CPL best-fit $\chi^2$ & UDF best-fit $\chi^2$ \\
\hline
Planck (w/o lensing) + DESI BAO &  $10968.66$ & $10970.40$\\
Planck (w/o lensing) + DESI BAO + DESY5 & $12608.00$ & $12609.99$\\
Planck (w lensing) + DESI BAO + DESY5 & $12617.73$ & $12618.69$ \\
\hline
\end{tabular}
\caption{Best-fit $\chi^2$ values for different combinations of data, when using either CPL or UDF. The best-fit values are very similar, even if CPL gives slightly better results. This slight improvement mainly comes from the difference in the CMB temperature power spectrum at large scales (see figure~\ref{fig:CMB}).\label{tab:best_fit}}
\end{table}

As can be seen, the best-fit $\chi^2$ between CPL and UDF is very similar for the three combinations of data we considered. The slightly better best-fit $\chi^2$ we obtain with CPL is due to the CMB ISW being lower in CPL predictions than in UDF. Conversely, we see that the preference for CPL decreases when adding CMB lensing reconstruction, showing that UDF fits CMB lensing data from Planck slightly better. Overall, those two models are very much equivalent in terms of goodness of fits to CMB (including lensing), BAO and supernovae data. 

\begin{figure}[htbp]
\centering
\includegraphics[width=.49\textwidth]{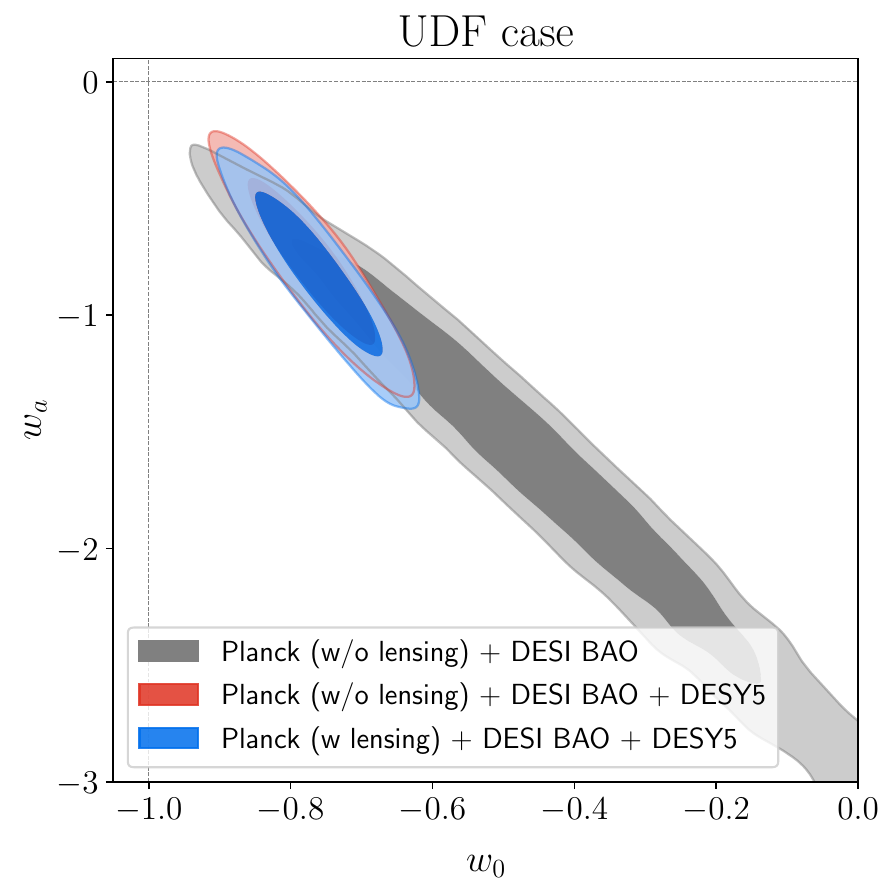}
\includegraphics[width=.49\textwidth]{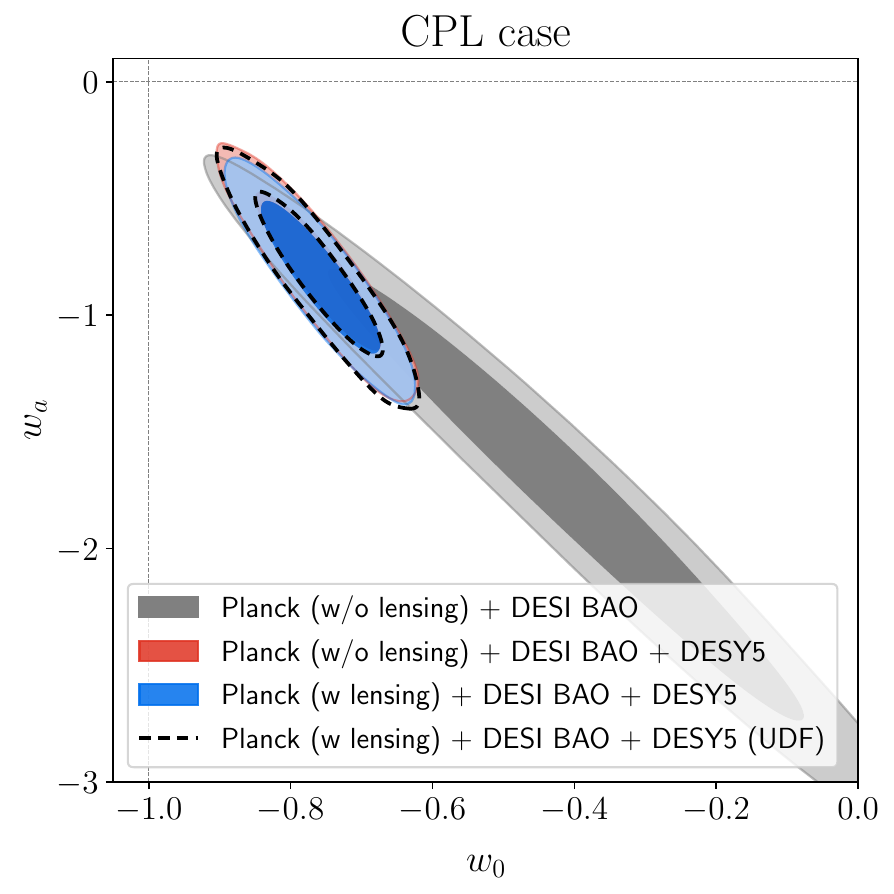}
\caption{Posterior distribution of $w_0$ and $w_a$ in the UDF model (left) and CPL (right), using Planck power spectra (with or without reconstructed lensing), DESI BAO DR2 and DESY5 supernovae. The contours we show represent the $68\%$ and $95\%$ confidence intervals. Using either UDF or CPL, we get almost the same contours, showing that with those sets of observations, the two models are very much equivalent. On the right plot, the dashed contour shows the result from the combination of CMB (including lensing), BAO, and supernovae observations using the UDF model, for comparison with the CPL contours.\label{fig:contours}}
\end{figure}

We also show on figure~\ref{fig:contours} the $68\%$ and $95\%$ posterior confidence intervals of $w_0$ and $w_a$ using the UDF model or CPL. We get very similar contours, showing that the two models indeed give similar predictions, even at different locations in the parameter space.

Our unified dark fluid, while phenomenologically simple, is somewhat unusual from a physical perspective. A single fluid with a negative and evolving equation of state but vanishing rest-frame sound speed does not occur in perfect-fluid descriptions, where adiabaticity would imply $c_s^2=w-\frac{1}{3\mathcal{H}}\frac{\dot{w}}{1+w}$~\cite{Hu:1998kj}. In our case, the combination $w<0$ and $c_s^2=0$ means that the fluid both drives accelerated expansion and clusters like cold matter. This double role is unexpected for a single perfect fluid, but it is not forbidden in the more general, non-adiabatic case. Our UDF model could indeed be a single non-adiabatic fluid, with evolving equation of state and null sound speed. If it is a single component, it could also correspond to a k-essence model~\cite{Armendariz-Picon:1999hyi,Armendariz-Picon:2000nqq}, in which one can choose the kinetic term and the potential of a scalar field, such that the equation of state and sound speed of the scalar field can be chosen independently. In these models, it is then possible to build a component with a negative equation of state and an arbitrarily low sound speed~\cite{Erickson:2001bq,Scherrer:2004au,Creminelli:2009mu,Kunz:2015oqa}. Alternatively, in multi-component or interacting dark sector models, the effective equation of state of the combined fluid can differ substantially from the equation of state of any individual constituent, while retaining an effective sound speed close to zero if this is the case for all components. In this sense, our UDF may be seen as a phenomenological proxy for a broader class of physically motivated models with low sound speed and negative pressure. 
In the latter picture, however, it should be noted that even though our UDF model can, in principle, reproduce the effect of a multi-component model, it may be difficult in practice to construct such a model with non-phantom subcomponents that would effectively match the combined equation of state shown in figure~\ref{fig:w_a}. Several interacting dark matter–dark energy models, for instance, only show moderate improvement over \lcdm\ when fitting DESI BAO data~\cite{Giare:2024smz,Ghedini:2024mdu,Sabogal:2024yha}. More complex models involving a time-varying coupling between dark matter and dark energy, such as those in~\cite{Petri:2025swg,Yang:2025uyv}, are required to fit those observations significantly better than \lcdm.

Importantly, despite its unconventional microphysics, our unified model is internally consistent, does not violate the Null Energy Condition, and passes many cosmological constraints. Its success therefore underscores that present data do not yet strongly constrain the clustering properties of the dark sector, and that background-level hints for phantom crossing should be interpreted with care.

\section{Future data: will UDF and CPL be distinguishable?}\label{sec:forecast}

As shown above, UDF and CPL make very similar predictions for the CMB, RSD, and lensing observables, and are identical for background probes by construction. It is therefore natural to ask whether upcoming stage IV CMB and galaxy surveys will have the statistical power to distinguish between them. In this section we perform a forecast based on CMB, \sixxtwo, BAO, and supernovae observables.

\paragraph{CMB}  
We generated fiducial power spectra assuming the CPL model as the true cosmology. Specifically, we produced mock \SO\ temperature spectra $C_\ell^{TT}$ for $\ell \in [40,3000]$ and polarization spectra $C_\ell^{EE}$ and $C_\ell^{TE}$ for $\ell \in [40,5000]$. Gaussian covariance matrices were computed using the baseline \SO\ noise curves~\cite{SimonsObservatory:2018koc} and  
\begin{multline}\label{eq:covariance}
    \mathrm{Cov}(C_\ell^{AB},C_{\ell'}^{CD}) = \frac{\delta_{\ell\ell'}}{(2\ell+1)\Delta\ell f_\mathrm{sky}}
    \Big[ (C_\ell^{AC}+N_\ell^{AC})(C_\ell^{BD}+N_\ell^{BD})
          \\
          +(C_\ell^{AD}+N_\ell^{AD})(C_\ell^{BC}+N_\ell^{BC}) \Big],
\end{multline}
where $A$, $B$, $C$ and $D$ are either the temperature or the E-mode polarization. We use multipole bins with constant width $\Delta\ell=10$ and a sky fraction $f_\textrm{sky}=0.4$. For low multipoles ($2\leq\ell\leq 40$), we added mock temperature spectra at Planck sensitivity, modelled as
\begin{align}\label{eq:Nl}
    N_\ell^{TT} = \sigma_T^2 \exp\!\left[\frac{\ell(\ell+1)\sigma_\mathrm{FWHM}^2}{8\ln 2}\right],
\end{align}
with $\sigma_T=23\,\mu\mathrm{K\,arcmin}$ and $\sigma_\mathrm{FWHM}=7\,\mathrm{arcmin}$, and adopted $f_\mathrm{sky}=0.7$. We did not include low-$\ell$ polarization spectra but imposed a Gaussian prior on the optical depth $\tau$ with $\sigma(\tau)=0.0073$~\cite{Planck18}.
\paragraph{\texorpdfstring{\boldmath $6\times2$\,pt}{6x2pt}}  
For the \sixxtwo\ observables, we constructed Euclid-like mock spectra with ten redshift bins, following the noise and covariance specifications of section~\ref{sec:6x2}. The CMB lensing noise was taken from \SO~\cite{SimonsObservatory:2018koc}. Since we only have linear predictions for the UDF, we conservatively restricted the forecast to the linear regime. We compared linear and non-linear CPL spectra (the latter from HMcode~\cite{Mead:2020vgs}) and imposed a cut at $\ell_\mathrm{max}$, defined separately for each probe, such that the non-linear corrections never exceed $5\%$~\cite{Loureiro:2018qva,Kou:2023gyc}. We adopted a linear galaxy bias per redshift bin and neglected weak-lensing systematics (intrinsic alignments, multiplicative shear) as well as cross-covariances between CMB and \sixxtwo, which have been shown to be subdominant~\cite{Kou:2025hvg}.

\paragraph{BAO and supernovae}  
For BAO and SN, we modified the DESI BAO DR2 and DESY5 likelihoods by replacing the data with CPL theoretical predictions evaluated at the fiducial cosmology.

\paragraph{Results}  
We then performed a $\chi^2$ minimization with Cobaya, varying 18 parameters (8 cosmological and 10 galaxy bias parameters). For CPL, the best-fit parameters we estimated yield $\chi^2_\mathrm{CPL}=0.21$, reflecting imperfect convergence of the minimizer. For UDF, we obtained $\chi^2_\mathrm{UDF}=6.77$, corresponding to $\Delta\chi^2=6.56$. Since the two models have the same number of parameters, this would also correspond to their difference in Akaike
Information Criterion (AIC)~\cite{Akaike74,Trotta17}. As a rule of thumb~\cite{burnham2002model,Liddle:2007fy,hallquist2014mixture}, an AIC difference smaller than $2$ suggests that the competing models are essentially indistinguishable, whereas a difference of $4$–$7$ points indicates substantially stronger support for the lower-AIC model, and a difference greater than $10$ points provides decisive support. In this context, a difference of $6.56$ implies a strong, though not decisive, preference for one model over the other, indicating that distinguishing between UDF and CPL will remain challenging even with stage IV surveys. Achieving stronger constraints will likely require incorporating (at least mildly) non-linear scales or additional cosmological probes.

\section{Model comparison in the non-linear regime}\label{sec:non_linear}

So far, our analysis has been restricted to the linear regime, leaving open the question whether the UDF model remains consistent once non-linear effects are taken into account. To partially address this, we study spherical collapse in the UDF framework and examine its impact on the redshift of collapse and on an approximate halo mass function. The aim here is not to derive a high-precision prediction suitable for direct comparison with observations, but rather to investigate qualitatively whether the UDF induces significant differences in small-scale structure formation. In particular, we compare the spherical collapse in the UDF model with a fiducial CPL case, where dark energy is assumed not to cluster on relevant scales. 

Spherical collapse in the context of clustering dark energy has already been considered in a number of studies~\cite{Abramo:2007iu,Creminelli:2009mu,Basse11,Batista:2017lwf,Heneka:2017ffk,Pace:2017qxv,Chang:2017vhs,Batista:2021uhb,Batista:2022ixz}, and the main differences here are to work with a single fluid, and to apply it to our fiducial model. We neglect the effect of photons and neutrinos (including massive neutrinos), and we work with a single fluid that includes baryons and the UDF which we assume to be comoving, as is mostly done in spherical collapse studies. 

\subsection{Linear growth}

In this section, we closely follow~\cite{Creminelli:2009mu}, which studied the spherical collapse in quintessence models
with zero speed of sound. In particular, we follow their use of Fermi coordinates within a relativistic framework. This means that we consider scales much smaller than the Hubble radius where spacetime is close to Minkowski. Fermi coordinates are then such that the deviation of the metric from Minkowski is suppressed by $H^2r^2$, where $r$ is the distance from a reference point for any time. In this context, we begin by examining the growth of structures in the linear regime. The density and velocity of the (baryons and UDF) fluid can be written as
\begin{align}
    \rho&=\bar{\rho}(1+\delta) \\
    \vec{v}&=H\vec{x}+\vec{u},
\end{align}
where $\vec{x}$ is the local coordinate, $\vec{u}$ is the perturbation to the Hubble flow velocity and an overbar indicates the background part of a quantity. Perturbing the Euler equation at linear order, \cite{Creminelli:2009mu}~showed
\begin{align}\label{eq:euler_perturbed}
    \vec{u}\,'+H\vec{u}+\frac{1}{a}\vec{\nabla}\delta\Phi=0,
\end{align}
using comoving gradient, where $\,'$ denotes derivatives with respect to physical time, and where $\delta\Phi$ is the perturbation of the Newtonian potential such that the Poisson equation gives
\begin{align}\label{eq:poisson}
    \frac{1}{a^2}\nabla^2\delta\Phi=4\pi G\bar{\rho}\delta.
\end{align}
Finally, perturbing the continuity equation, \cite{Creminelli:2009mu}~obtained
\begin{align}\label{eq:continuity_perturbed}
    \delta'-3Hw\delta+(1+w)\frac{1}{a}\vec{\nabla}\cdot\vec{u}=0,
\end{align}
where $w$ denotes the (time varying) equation of state of the (total) fluid. This equation is different from~\ref{eq:perturbation_density} because on small scales (in the local Fermi patch), relativistic metric terms as well as terms proportional to $\mathcal{H}v/k$ are suppressed by $(\mathcal{H}/k)^2$ and can be neglected. The latter terms originate from the adiabatic sound speed $c_a^2=\frac{\dot{p}}{\dot{\rho}}=w+\frac{\dot{w}\rho}{\dot{\rho}}$, which even for null rest frame sound speed is non-zero in general, and contributes to the pressure perturbations such that~\cite{Bean:2003fb}
\begin{align}
    \delta p=\hat{c}_{s}^2\delta\rho-\left(\hat{c}_{s}^2-c_a^2\right)\dot{\rho}\frac{\theta}{k^2}.
\end{align}
Using $\hat{c}_s^2=0$ and replacing $\dot{\rho}$ with the continuity equation, we get
\begin{align}
    \delta p = (\dot{w}-3\mathcal{H}(1+w)w)\frac{\rho\theta}{k^2}.
\end{align}
This vanishes in the synchronous gauge as $\theta=0$ (neglecting baryon velocity), whereas in the Newtonian gauge, using equation~\ref{eq:velocity_density} and converting (conformal) time derivatives to scale factor derivatives, we get
\begin{align}
    \delta p = (3(1+w)w-a\frac{dw}{da})f_\textrm{eff}\rho\delta\frac{\mathcal{H}^2}{k^2},
\end{align}
making explicit the suppression by $(\mathcal{H}/k)^2$. More generally, on scales well inside the Hubble radius, the choice of gauge does not affect this conclusion, and pressure perturbations can safely be neglected.

We now differentiate equation~\ref{eq:continuity_perturbed} with respect to time, replace the term in $\vec{\nabla}\cdot\vec{u}\,'$ taking the divergence of equation~\ref{eq:euler_perturbed}, and use equation~\ref{eq:poisson} to replace $\nabla^2\delta\Phi$. We finally change time derivatives to scale factor derivatives, and end up with

\begin{multline}
\frac{d^2\delta}{da^2}+\left[\frac{1}{H}\frac{dH}{da}+\frac{3}{a}(1-w)-\frac{1}{1+w}\frac{dw}{da}\right]\frac{d\delta}{da} \\
-\left[\frac{3}{a(1+w)}\frac{dw}{da}+\frac{6w}{a^2}+\frac{3w}{aH}\frac{dH}{da}+\frac{4\pi G}{a^2H^2}(1+w)\bar{\rho}\right]\delta=0.
\end{multline}
This equation allows us to compute the evolution of the growth parameter $D$ as a function of scale factor, such that in linear theory, $\delta(a)=\delta_i\frac{D(a)}{D(a_i)}$. Integrating this differential equation, we found very good agreement with the growth parameter computed in CAMB.

\subsection{Spherical collapse}
We now turn to the spherical collapse of a homogeneous overdensity of radius $R$. For a fluid with null sound speed, the evolution of $R$ is given by~\cite{Creminelli:2009mu}
\begin{align}\label{eq:Friedmann_local}
    \frac{R''}{R}=-\frac{4\pi G}{3}(\rho+3\bar{p})
\end{align}
where $\rho$ includes perturbations, but the pressure $\bar{p}=p$ has no perturbation in our case, as previously mentioned. To assess the impact of our model, we compare the spherical collapse of the UDF with that of CPL. In the CPL case, we assume that dark energy is unperturbed on the relevant scales, so that $\rho_\textrm{CPL}=\bar{\rho}_\textrm{CPL}$. This treatment differs slightly from the PPF approach adopted in the previous sections, but remains consistent, since CPL dark energy is nearly unclustered on the subhorizon scales considered here. We then use $\rho+3\bar{p}=\bar{\rho}(1+\Delta+3w)$, where $\Delta$ is the (non-linear) overdensity of the fluid, by contrast with $\delta$ which is the overdensity in the linear theory. Replacing time derivatives by scale factor derivatives, switching $R$ to $y=R/R_i$, where $R_i$ is the initial radius of the sphere, and using $\Omega(a)=\frac{8\pi G}{3H_0^2}\bar{\rho}(a)$, we obtain
\begin{align}
    \frac{d^2y}{da^2}+\left(\frac{1}{a}+\frac{1}{H}\frac{dH}{da}\right)\frac{dy}{da}+\frac{H_0^2}{2a^2H^2}\Omega(a)(1+\Delta+3w)y=0.
\end{align}
We also need to compute the evolution of the overdensity of the sphere $\Delta$. To do so, we consider the continuity equation of the fluid inside the sphere
\begin{align}
    \rho'+3\frac{R'}{R}
(\rho+\bar{p})=0.
\end{align}
Using $\rho=\bar{\rho}(1+\Delta)$ and $\bar{p}=\bar{\rho}w$,
\begin{align}
    (1+\Delta)\frac{d\bar{\rho}}{da}+\frac{d\Delta}{da}\bar{\rho}+\frac{3}{R}\frac{dR}{da}\bar{\rho}(1+\Delta+w)=0.
\end{align}
We then replace $\frac{d\bar{\rho}}{da}$ by its background value and switch $R$ to $y$, such that
\begin{align}
    \frac{d\Delta}{da}+\left(\frac{3}{y}\frac{dy}{da}-\frac{3}{a}(1+w)\right)\Delta=3(1+w)\left(\frac{1}{a}-\frac{1}{y}\frac{dy}{da}\right).
\end{align}
We hence end up with two coupled differential equations and need initial conditions for $y(a_i)$, $\frac{dy}{da}(a_i)$ and $\Delta(a_i)$. By definition, we have $y(a_i)=1$, we make $\Delta(a_i)=\Delta_i$ an input parameter, and only need an expression for $\frac{dy}{da}(a_i)$. We use an initial scale factor $a_i=10^{-5}$ such that $w(a_i)\approx0$ and matter is conserved initially. As a result, $\frac{4\pi}{3}R^3\bar{\rho}(1+\delta)=\frac{4\pi}{3}R^3\bar{\rho}(1+\delta_i\frac{D(a)}{D(a_i)})$ is constant. Differentiating with respect to scale factor yields
\begin{align}
    \frac{dR}{da}=\frac{R}{a}-\frac{R}{3}\frac{\delta_i}{D(a_i)+\delta_iD(a)}\frac{dD}{da}.
\end{align}
We hence have $\frac{dy}{da}(a_i)=\frac{1}{a_i}\left[1-\frac{1}{3}\frac{\delta_i}{1+\delta_i}\frac{d\ln{D}}{d\ln{a}}(a_i)\right]$, with $\delta_i=\Delta_i$ at the initial time.

\begin{figure}[htbp]
\centering
\includegraphics[width=\textwidth]{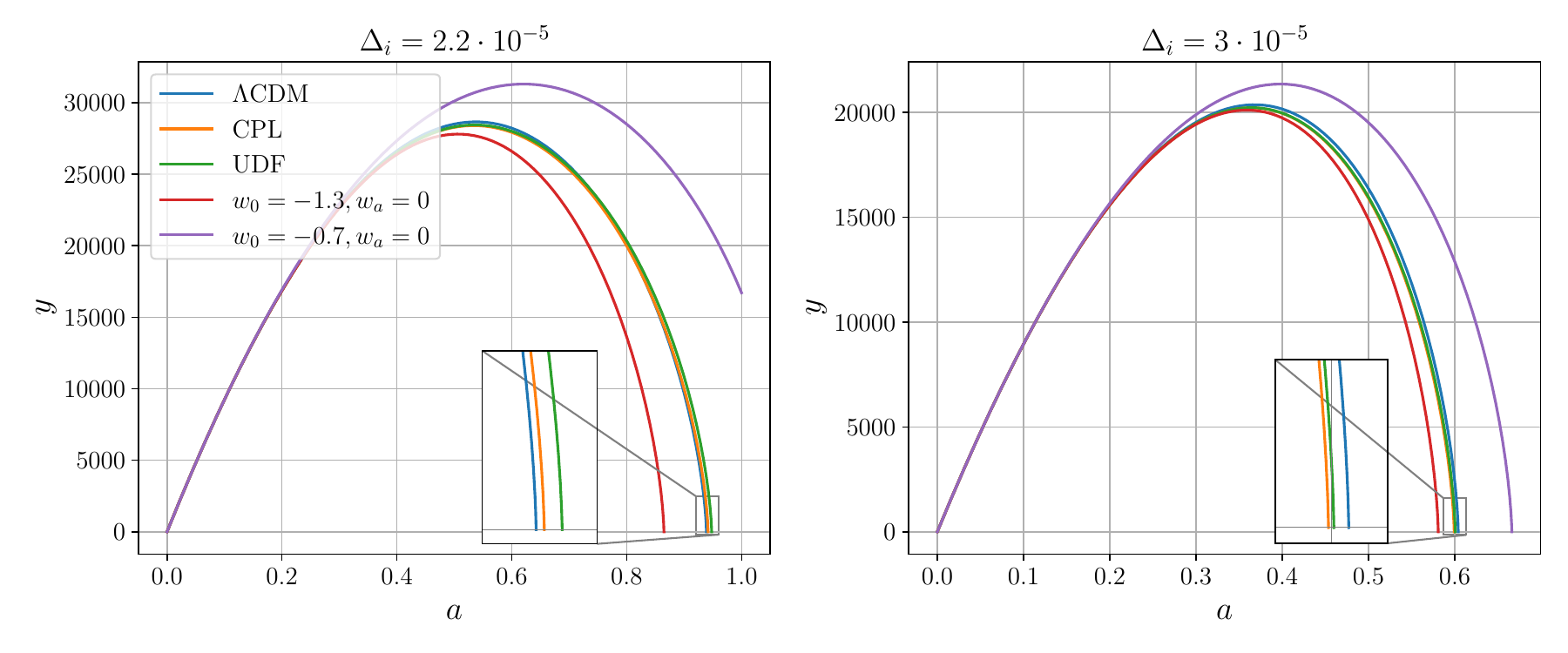}
\caption{Evolution of the normalized radius of a spherical overdensity as a function of scale factor, for initial perturbations $\Delta_i=2.2\cdot10^{-5}$ (left panel) or $\Delta_i=3\cdot10^{-5}$ (right panel). The models considered are \lcdm, our fiducial UDF model ($w_0=-0.75$, $w_a=-0.83$), two additional UDF cases ($w_0=-1.3$, $w_a=0$ and $w_0=-0.7$, $w_a=0$), and the CPL case. The latter refers to the case where matter and dark energy are not considered as unified, and only matter perturbations are taken into account, dark energy only having an impact through its effect on the background expansion of the Universe.\label{fig:y_vs_a}}
\end{figure}

Figure~\ref{fig:y_vs_a} shows the evolution of the normalized radius $y$ for two choices of initial overdensity, $\Delta_i=2.2\cdot10^{-5}$ and $\Delta_i=3\cdot10^{-5}$, across different models. Besides \lcdm, we display the fiducial UDF ($w_0=-0.75$, $w_a=-0.83$), two additional null sound speed UDF cases ($w_0=-1.3$, $w_a=0$ and $w_0=-0.7$, $w_a=0$) where we keep present density parameters ($\Omega_ch^2$, $\Omega_bh^2$) as well as $H_0$ fixed, and the CPL case with unperturbed dark energy. In the latter case, CPL dark energy has an impact on the spherical collapse from its effect on the background expansion only. When we model it, we do not use the unified framework and work only with the matter density, velocity, etc., taking dark energy contributions at the background solely. On small scales, standard dark energy clusters much less than matter, as assumed by the PPF approach, which is the reason why we neglect dark energy perturbations in the CPL case. In particular, when we refer to CPL, we use the same fiducial values as in the rest of the paper, i.e. $w_0=-0.75$ and $w_a=-0.83$. In \lcdm, dark energy is unperturbed ($w=-1$), and we checked that we get exactly the same result when using either a unified (matter and dark energy combined) or non-unified fluid.

Figure~\ref{fig:y_vs_a} shows that, for our choice of fixed present-day parameters ($\Omega_bh^2$, $\Omega_ch^2$ and $H_0$), a more negative equation of state of the unified dark fluid leads to an earlier collapse. This behaviour can be understood from the term $\rho+3\bar{p}=\bar{\rho}(1+\Delta+3w)$ that enters the collapse equation (\ref{eq:Friedmann_local}). At fixed background density, a smaller $w$ would strengthen the opposing pressure term and thus delay collapse. However, since we keep the present-day densities fixed, a more negative $w$ also implies a lower background density at earlier times, which weakens the opposing effect and makes the collapse occur earlier. In our configuration this background-density effect dominates over the instantaneous pressure term, leading to the trend seen in Figure~\ref{fig:y_vs_a}. Our results agree with findings from~\cite{Creminelli:2009mu} in the context of clustering dark energy.

It is interesting to compare the UDF and CPL cases as they share the same background expansion but differ at the perturbation level. Although the UDF is fully unified in our modelling, we can, for the sake of interpretation, conceptually separate it into a cold matter-like part (with current density $\Omega_ch^2+\Omega_bh^2$ and null equation of state) and a dark energy-like part. This decomposition is not used in the equations but helps to describe the origin of the differences between UDF and CPL, which arise precisely from how the “dark energy part” of the UDF behaves. In the case where $\Delta_i=3\cdot10^{-5}$ (right panel of figure~\ref{fig:y_vs_a}), the equation of state of the dark energy part of the fluid has always been below $-1$ between $a_i=10^{-5}$ and the collapse around $a\approx0.6$. As a result, it is expected that, in both CPL and UDF cases, the dark energy density will be lower than in \lcdm, which explains why the collapse happens earlier than in \lcdm\ (there has been less dark energy to oppose the collapse). It can also be seen that the collapse happens later in UDF than in CPL. As explained in~\cite{Abramo:2007iu,Creminelli:2009mu}, for dark energy with equation of state $w_{de}$ below $-1$, positive matter perturbations $\delta_m$ are associated with negative dark energy perturbations $\delta_{de}$ because $\delta_{de}\sim(1+w_{de})\delta_m$. These negative perturbations counteract collapse, delaying it relative to the CPL case where dark energy is considered unperturbed. On the left panel, when $\Delta_i=2.2\cdot10^{-5}$, the collapse happens later, such that for our fiducial values of $w_0$ and $w_a$ in UDF and CPL, it happens when the dark energy equation of state has become greater than $-1$. It turns out that in this fiducial set-up, the dark energy has had an equation of state greater than $-1$ for enough time for the dark energy to slow down the collapse and make it happen later than in \lcdm. However, while dark energy perturbations with $w>-1$ usually tend to enhance the collapse, we still find UDF to collapse later than CPL. The reason is that for most of the evolution, the equation of state was below $-1$, so the suppressive effect of negative perturbations dominates, and the subsequent positive perturbations have not had sufficient time to compensate.

\begin{figure}[htbp]
\centering
\includegraphics[width=\textwidth]{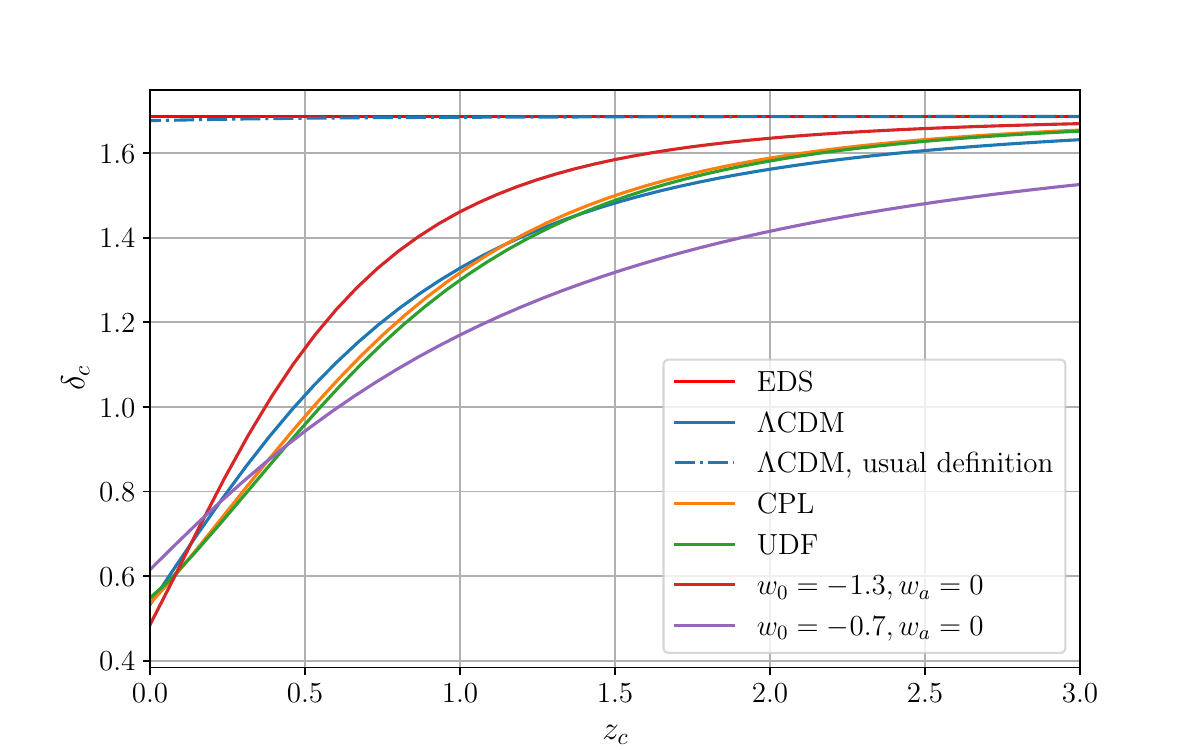}
\caption{Critical density contrast $\delta_c$ as a function of the redshift of collapse $z_c$ for the same models as in figure~\ref{fig:y_vs_a}. The Einstein-de Sitter (EDS) limit is shown for reference. In the unified-fluid case, $\delta_c$ is defined for the total density, which explains the unusual low-redshift behaviour where dark energy contributes significantly. For the CPL case, we compute the matter-only critical density contrast and rescale it by $\bar{\rho}_m(z_c)/(\bar{\rho}_m(z_c)+\bar{\rho}_{de}(z_c))$ for comparison. We also show, as a dashed line, the critical density contrast we obtain in \lcdm\ using the standard definition, which excludes dark energy from the denominator. This highlights the very significant effect of our redefinition.\label{fig:delta_c_vs_z_c}}
\end{figure}

\subsection{Critical density contrast}
We next compute the critical density contrast $\delta_c$, defined as the linear overdensity at the collapse redshift. In practice, this is obtained by evolving the initial overdensity with linear perturbation theory until the time of collapse. This quantity will then be useful to compute halo mass functions, as the idea is to evolve an initial density field linearly and then to compare the overdensities with $\delta_c$ to estimate whether a halo should have formed. Figure~\ref{fig:delta_c_vs_z_c} shows the evolution of $\delta_c$ as a function of the redshift of collapse $z_c$ for the same models as in figure~\ref{fig:y_vs_a}, with the Einstein-de Sitter (EDS) case included for reference.
Since we work in a unified-fluid framework, our $\delta_c$ is defined with respect to the total fluid, which includes the contribution that would correspond to dark energy in a non-unified treatment. This explains the somewhat unusual behaviour at low redshift, where dark energy contributes significantly to the total density. For the CPL case, where we treat dark energy as unperturbed, we instead compute the usual matter-only critical density contrast and rescale it by the factor $\bar{\rho}_m(z_c)/(\bar{\rho}_m(z_c)+\bar{\rho}_{de}(z_c))$, so that it can be compared consistently with the unified definition.

As expected, all models converge to the EDS value of $\delta_c$ at high redshift, where dark energy becomes negligible. At all redshifts, the fiducial UDF and CPL models follow nearly identical trajectories, and both differ more from \lcdm\ than from each other. This indicates that the main impact on $\delta_c$ arises from differences in the background expansion rather than from dark energy perturbations.

\subsection{Halo mass function
}
Finally, we compute the halo mass function using the Press-Schechter formalism~\cite{1974ApJ...187..425P,1991ApJ...379..440B}, such that
\begin{align}\label{eq:Press_schechter}
    \frac{dn}{dM}(M,z)=-\sqrt{\frac{2}{\pi}}\frac{\bar{\rho}}{M^2}\frac{\delta_c(z)}{D(z)\sigma_R}\frac{d\log{\sigma_R}}{d\log{M}}\exp\left(-\frac{\delta_c^2(z)}{2D^2(z)\sigma_R^2}\right),
\end{align}
where $D(z)$ is the growth factor normalized at $z=0$, and $\sigma_R$ is the variance of the smoothed density field,
\begin{align}
    \sigma_R^2=\frac{1}{2\pi^2}\int_0^\infty dk k^2\lvert W(kR)\lvert^2P(k),
\end{align}
with $P(k)$ the total power spectrum and $W(kR)=3(\sin(kR)-kR\cos(kR))/(kR)^3$ the Fourier transform of a top-hat filter. We compute $\sigma_R$ using our modified version of CAMB.

A subtle point concerns how to define the halo mass in a unified model. In studies of clustering dark energy, the mass is usually taken to include matter and dark energy perturbations, but not the homogeneous dark-energy background~\cite{Creminelli:2009mu,Batista:2017lwf,Heneka:2017ffk}. This choice is somewhat arbitrary, but reasonable when matter and dark energy are treated as separate fluids. In our single fluid UDF description, however, such a split is not possible. We therefore define the halo mass as
\begin{align}
    M(a)=\frac{4\pi}{3}R^3\bar{\rho}(a).
\end{align}
Since $\bar{\rho}(a)$ does not evolve as $a^{-3}$, the halo mass is not conserved in general, as in most other mass definitions.

Because of our mass definition, our halo mass function does not exactly match the usual estimates in the \lcdm\ case. However, within a given model, the collapse redshift is unaffected by how the total fluid is decomposed, provided that all components are comoving. In other words, one could equivalently follow the collapse of the total fluid (as we do), of its matter part, or of any comoving subcomponent, as they all collapse simultaneously. In particular, the ratio $\delta_c(z)/(D(z)\sigma_R)$ remains the same regardless of which comoving component is used to describe the collapse. What changes is simply how a given physical halo is assigned a mass under different conventions. It is then easy enough to convert the halo mass function in different definitions by rescaling the mass. To facilitate comparison, we hence give the relation between halo mass functions in our definition or in terms of the matter-only mass definition,
\begin{align}\label{eq:convert_dndm}
    \frac{dn}{d\log{M}}\left(M,z\right)=\frac{dn}{d\log{M_m}}\left(M_m=\frac{\bar{\rho}_m(z)}{\bar{\rho}(z)}M,z\right).
\end{align}
where $M_m$ denotes the matter-only halo mass. In particular, we use this relation to convert the matter-only halo mass function we computed in the CPL case, and express it in the total fluid convention.

\begin{figure}[htbp]
\centering
\includegraphics[width=\textwidth]{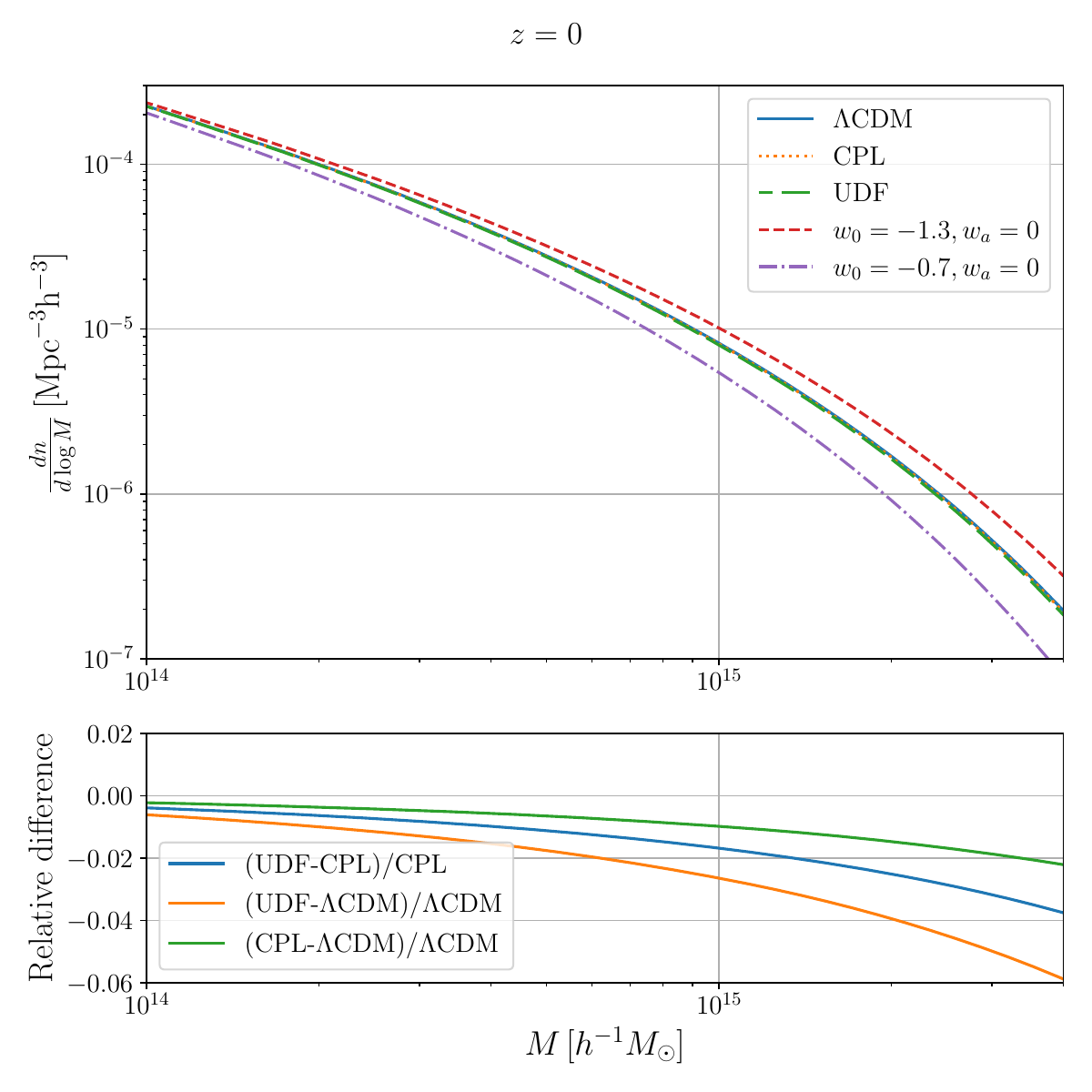}
\caption{Number of halos per logarithmic mass range as a function of halo mass (top panel) and relative differences between \lcdm, CPL and UDF (bottom panel) at $z=0$.\label{fig:dndm_z0}}
\end{figure}

\begin{figure}[htbp]
\centering
\includegraphics[width=\textwidth]{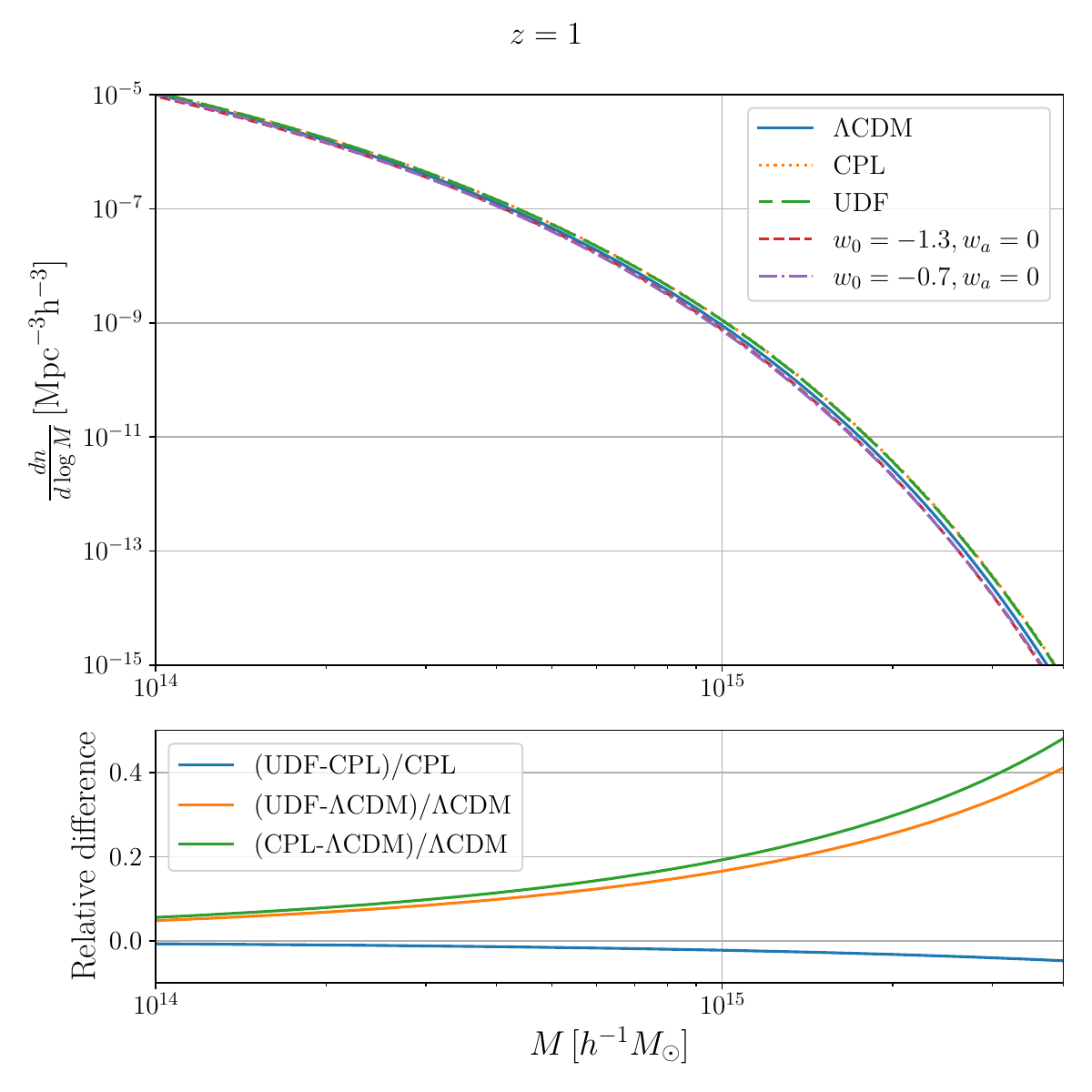}
\caption{Same as figure~\ref{fig:dndm_z0} at $z=1$.\label{fig:dndm_z1}}
\end{figure}

Figures~\ref{fig:dndm_z0} and~\ref{fig:dndm_z1} show the halo mass functions at $z=0$ and $z=1$, respectively. At $z=0$, we recover the trend found in~\cite{Creminelli:2009mu}: lowering the fluid's equation of state increases the number of halos, since collapse occurs earlier and more structures can form. The fiducial UDF and CPL models both yield very similar results (close to \lcdm\ as well), with differences of only a few percent at the high-mass end. Notably, UDF tends to produce slightly fewer halos than CPL, consistent with the delayed collapse caused by the clustering of the dark energy part.

Looking at $z=1$ (figure~\ref{fig:dndm_z1}), the picture changes because two competing effects are at play. Firstly, as we mentioned previously, at fixed halo mass, a higher equation of state of the unified dark fluid suppresses clustering, delaying collapse and eventually reducing the halo abundance. The second effect is that, for a given initial overdensity, the resulting halo is more massive when the fluid’s equation of state is higher, since the fluid’s background density is also higher and thus contributes more to the halo mass. This effect tends to shift the halo mass functions towards the higher mass end. Because those two effects go in opposite directions, there is no clear ordering of the halo mass functions any more at $z=1$. This effect did not happen at $z=0$ because given the parameters that we keep fixed (density parameters and $H_0$), the density of the fluid at $z=0$ is the same in all the models we considered. Finally, we checked using equation~\ref{eq:convert_dndm} that the second effect we mentioned does not appear if we do not include the dark energy part in the halo mass, which confirms our interpretation.

At $z=1$, we also find a much bigger difference between \lcdm\ and either CPL or UDF halo mass functions, which seems to be primarily due to the differences in the background expansion of the Universe as differences between UDF and CPL are rather small. We also checked that this difference is not related to our mass definition. It is likely that the very small differences we find at $z=0$ is due to the fact that the equation of state of the dark energy part crosses $-1$, so that halo formation is enhanced initially, and hindered later, whereas at $z=1$, the dark energy equation of state has not crossed $w=-1$ yet and halo formation is only enhanced. When comparing UDF and CPL, we still find very small differences, showing that our UDF model would not be obviously ruled out by non-linear observations, either at redshift $z=0$ or $z=1$. We note however that this conclusion may change in case one used a more motivated mass definition, looking more carefully at what fraction of the fluid is actually clustering, which we leave for future work.

\section{Conclusion}

We proposed a unified dark matter–dark energy model with vanishing sound speed. This UDF scenario reproduces DESI BAO DR2, Planck, and DESY5 data almost as well as CPL, with a $\chi^2$ difference below unity. By construction, its background expansion matches CPL, while the perturbations evolve differently. We assessed the consequences for a wide range of observables, namely CMB spectra, the \sixxtwo\ analysis, and RSD, and consistently found only modest deviations from CPL. Forecasts for stage IV experiments indicate that they may achieve a statistical preference between the two models. However, decisive evidence will remain challenging unless the true cosmology departs more strongly from \lcdm\ than our fiducial case, or unless additional probes or smaller (mildly non-linear) scales are included. Extending our work to the non-linear regime through the study of the spherical collapse, we also find that the UDF gives similar predictions from CPL, even though more work would be necessary to identify clearly how UDF would contribute to halo masses. 

From a theoretical standpoint, the UDF can be interpreted in several ways: as a single fluid with evolving equation of state and null sound speed, as a clustering quintessence field such as k-essence, or as a collection of (interacting or not) components with vanishing sound speeds.

Our model performs comparably to CPL with current data and yields similar forecasts for next-generation surveys. It provides a physical description of perturbations while ensuring consistency between background and perturbation levels. As such, it offers a way of interpreting recent DESI and supernovae results. In particular, it accommodates the apparent hints of phantom-crossing dark energy and NEC violation without introducing instabilities or pathologies. More generally, our results highlight that deviations from \lcdm\ can be interpreted in multiple physically consistent ways, emphasizing the diversity of viable cosmological models.
\acknowledgments
We are supported by UK STFC grant ST/X001040/1.

\bibliographystyle{JHEP}
\bibliography{biblio}

\end{document}